\documentclass[cits]{PoS}

\newcommand{\aj}{AJ}
\newcommand{\apj}{ApJ}
\newcommand{\apjl}{ApJ}
\newcommand{\apjs}{ApJS}
\newcommand{\aap}{A\&A}

\newcommand{\mnras}{MNRAS}

\newcommand{\prd}{Phys. Rev. D}

\title{The galaxy cluster concentration-mass relation in dark energy cosmologies}

\ShortTitle{{\rm{c-M}} relation in dark energy cosmologies}

\author{\speaker{Cristiano De Boni}\thanks{In collaboration with Klaus Dolag, Stefano Ettori, Lauro Moscardini and Valeria Pettorino.}\\
        Dipartimento di Fisica e Astronomia, Universit\`a di Bologna \\
        viale Berti Pichat 6/2, I-40127 Bologna, Italy \\ 
        E-mail: \email{cristiano.deboni@unibo.it}}

\abstract{We use numerical simulations of different dark energy cosmologies to investigate the concentration-mass ($c-M$) relation in galaxy clusters. In particular, we consider a reference $\Lambda$ cold dark matter ($\Lambda$CDM) model, two models with dynamical dark energy, viewed as a quintessence scalar field [using a Ratra and Peebles (RP) and a supergravity (SUGRA) potential form], and two extended quintessence models, one with positive and one with negative coupling (EQp and EQn respectively), where the quintessence scalar field interacts non-minimally with gravity (scalar-tensor theories). All the models are normalized in order to match CMB data from {\it{Wilkinson Microwave Anisotropy Probe}} 3 (WMAP3). For each model, we have performed numerical simulations in a cosmological box of $(300 \ {\rm{Mpc}} \ h^{-1})^{3}$. We fit the dark matter profile with a Navarro-Frenk-White (NFW) profile, and recover the concentration of each halo. We consider both the complete catalog of clusters and groups and a subsample of relaxed objects. The $c-M$ relation of our reference $\Lambda$CDM model is in good agreement with the results in literature, and relaxed objects have a higher normalization and a shallower slope with respect to the complete sample. For the different dark energy models, we find that for $\Lambda$CDM, RP and SUGRA the normalization of the $c-M$ relation is linked to the growth factor $D_{+}$ and the power spectrum normalization $\sigma_{8}$, with models having a higher value of $\sigma_{8} D_{+}$ having also a higher normalization. This simple scheme is no longer valid for EQp and EQn because in these models it is present a time dependent effective gravitational interaction, whose redshift evolution depends on the sign of the coupling. This leads to a decrease (increase) of the expected normalization in the EQp (EQn) model. This result shows a direct manifestation of the coupling between gravity and the quintessence scalar field characterizing EQ models that cannot be seen at the background level but can be investigated in the non-linear regime.}

\FullConference{Proceedings of the Corfu Summer Institute 2012 "School and Workshops on Elementary Particle Physics and Gravity"\\
		September 8-27, 2012\\
		Corfu, Greece}

\begin{document}

\section{Introduction}

\noindent Over the last decade great observational evidence (\cite{1998AJ....116.1009R,1999ApJ...517..565P,2011ApJS..192...14J,2009ApJ...692.1060V}) has shown that at the present time the Universe is expanding at an accelerated rate. This fact can be attributed to a component with negative pressure, which is usually referred to as dark energy, that today accounts for about 3/4 of the entire energy budget of the Universe. The simplest form of dark energy is a cosmological constant term $\Lambda$ in Einstein's equation, within the so-called $\Lambda$ cold dark matter ($\Lambda$CDM) cosmologies. Though in good agreement with observations, a cosmological constant is theoretically difficult to understand in view of the fine-tuning and coincidence problems. A valid alternative consists in a dynamical dark energy contribution that changes in time and space, often associated to a scalar field (`quintessence') evolving in a suitable potential (\cite{1988NuPhB.302..645W,1988PhRvD..37.3406R}). Dynamical dark energy allows for appealing scenarios in which the scalar field is the mediator of a fifth
force, either within scalar-tensor theories or in interacting scenarios (\cite{1995A&A...301..321W,2000PhRvD..62d3511A, 2000PhRvL..85.2236B,2008PhRvD..77j3003P,2008PhLB..663..160M} and references therein). 
In view of future observations, it is of fundamental interest to investigate whether dark energy leaves some imprints in structure formation, giving a practical way to distinguish among different cosmologies (\cite{2007PhRvD..76f4004H,2010MNRAS.403.1684B,2010ApJ...712L.179Z,2011MNRAS.411.1077B,2011MNRAS.412L...1B,2010PhRvD..81f3525W}).

The internal properties of dark matter halos are known to reflect their formation history and thus the evolution of the background cosmology. \cite{1996ApJ...462..563N} (hereafter NFW) found that the dark matter profile of a halo can be characterized by a scale radius, which is linked to the virial radius through the concentration of the object. The concentration of a dark matter halo is related to the mean density of the universe at the halo formation time.

Because of the hierarchical nature of structure formation and the fact that collapsed objects retain information on the background average matter density at the time of their formation (\cite{1996ApJ...462..563N}), concentration and mass of a dark matter halo are related. Since low-mass objects form earlier than high-mass ones, and since in the past the background average matter density was higher, low-mass halos are expected to have a higher concentration compared to high-mass ones. These expectations have been confirmed by the results of $N$-body numerical simulations which find, at $z=0$, a concentration-mass relation $c(M) \propto M^{\alpha}$, with $\alpha \sim -0.1$ (\cite{2004A&A...416..853D,2008MNRAS.387..536G,2010ApJ...712L.179Z}), with a log-normal scatter ranging from $0.15$ for relaxed systems to $0.30$ for disturbed ones (\cite{2000ApJ...535...30J}).

$N$-body simulations have been carried out by several authors in order to study the $c-M$ relation in dark matter halos with sizes of galaxy groups and clusters. \cite{2004A&A...416..853D} performed simulations with different cosmological models in order to verify the effects of dark energy dynamics. They found that the halo concentration depends on the dark energy equation of state through the linear growth factor at the cluster formation redshift, $D_{+}(z_{coll})$. \cite{2007MNRAS.381.1450N} also noted that non-relaxed objects have a lower concentration and a higher scatter with respect to relaxed ones. \cite{2008MNRAS.391.1940M} made a comparison between concentrations in the {\it{Wilkinson Microwave Anisotropy Probe}} 1 (WMAP1), WMAP3 and WMAP5 cosmologies in order to study the effects of different cosmological parameters (in particular the power spectrum normalization $\sigma_{8}$) on the $c-M$ relation. 

Since the concentration of a halo is linked to the background density of the universe at the time it collapsed, and since different dark energy models predict different evolutions of the cosmological background, it is interesting to investigate the impact of dark energy on the $c-M$ relation. Moreover, since some dark energy models can also affect the linear and non linear evolution of the density fluctuations, leaving some imprints in collapsed structure, one can think about using the $c-M$ relation as a cosmological probe, orthogonal to others that are commonly used. $N$-body cosmological simulations of extended quintessence models, including the effects on the $c-M$ relation, were presented in \cite{2011ApJ...728..109L}.

\section{Dark energy models} \label{models}

\noindent We consider the same cosmological models discussed in \cite{2011MNRAS.415.2758D}. Here we recall only the main features of the different models, and refer to \cite{2011MNRAS.415.2758D} for more details.

As a reference model we use the concordance $\Lambda$CDM model. This model is characterized by the presence of a dark energy component given by a cosmological constant $\Lambda$, with equation of state $w_{\Lambda}=-1$. 

The second case is a model with dynamical dark energy, given by a quintessence scalar field $\phi$ with an equation of state evolving with redshift, $w=w(z)$ (\cite{1988NuPhB.302..645W,1988PhRvD..37.3406R}). As in \cite{2011MNRAS.415.2758D}, as potentials for minimally coupled quintessence models, we consider an inverse power-law potential

\begin{equation}
\label{rp_potential}
V(\phi)=\frac{M^{4+\alpha}}{\phi^{\alpha}} \ ,
\end{equation}

\noindent the so called RP potential (\cite{1988PhRvD..37.3406R}), as well as its generalization suggested by supergravity arguments (\cite{1999PhLB..468...40B}), known as SUGRA potential, given by

\begin{equation}
\label{sugra_potential}
V(\phi)=\frac{M^{4+\alpha}}{\phi^{\alpha}}\exp(4\pi G \phi^2) \ ,
\end{equation}

\noindent where in both cases $M$ and $\alpha \ge 0$ are free parameters (see Table \ref{tab} for details).

The third possibility we consider is the case in which $\phi$ interacts non minimally with gravity (\cite{1988NuPhB.302..645W,2000PhRvL..85.2236B}). In particular we refer to the extended quintessence (EQ) models described in \cite{2000PhRvD..61b3507P}, \cite{2005JCAP...12..003P} and \cite{2008PhRvD..77j3003P}. The parameter $\xi$ represents the "strength" of the coupling (see Table \ref{tab} for details). In particular we consider here a model with positive coupling $\xi > 0$ (EQp) and one with negative $\xi < 0$ (EQn). For an extensive linear treatment of EQ models we refer to
\cite{2008PhRvD..77j3003P}. Here we only recall for convenience that
EQ models behave like minimally coupled quintessence theories in
which, however, a time dependent effective gravitational interaction
is present. In particular, in the Newtonian limit, the gravitational
parameter is redefined as

\begin{equation} 
\label{EQ_Gtilde_def} 
\tilde{G} = \frac{2 [ F + 2(\partial F/\partial \phi)^2]}{[ 2 F+3(\partial F/\partial \phi)^2]} \frac{1}{8 \pi F} \ .
\end{equation} 

\noindent Here the coupling $F(\phi)$ is chosen to be

\begin{equation}
\label{non-minimal coupling}
F(\phi)=\frac{1}{\kappa}+\xi (\phi^2 - {\phi_{0}^2}) \ ,
\end{equation}

\noindent with $\kappa=8\pi G_\ast$, where $G_{\ast}$ represents the ``bare'' gravitational constant (\cite{2001PhRvD..63f3504E}).

\begin{figure}
\begin{center}
\hspace*{-0.3in}
\includegraphics[width=0.5\textwidth]{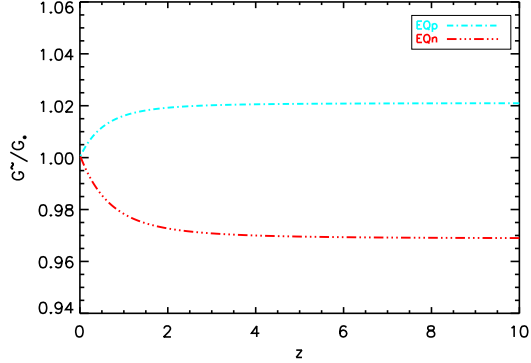}
\end{center}
\caption{Correction to the gravity constant for the two extended quintessence models, EQp (cyan) and EQn (red), as expressed in equation (2.5). Note that the corrections are only within the percent level.}
\label{dG_z}
\end{figure}

\noindent For small values of the coupling, that is to say $\xi \ll 1$, the latter expression becomes

\begin{equation}  
\frac{\tilde{G}}{\ G_\ast} \sim 1 - 8 \pi G_\ast \xi (\phi^2 - {\phi_{0}^2}) \ ,
\label{dG}
\end{equation}

\noindent which manifestly depends on the sign of the coupling $\xi$. We note that, since the derivative of the RP potential in equation (\ref{rp_potential}) with respect to $\phi$ is $\partial V(\phi)/ \partial \phi < 0$, we have $\phi^2 < {\phi_{0}^2}$. This leads to the behaviour of ${\tilde{G}}/{G_\ast}$ shown in Fig. \ref{dG_z}. Note that the corrections are only within the percent level.

\section{Numerical simulations}

\noindent In order to study the formation and evolution of large scale structures in these different cosmological scenarios we use $N$-body simulations performed with the {\small{GADGET-3}} code (\cite{2001ApJ...549..681S,2005MNRAS.364.1105S}).
For each model, we simulated a cosmological box of size $(300 \ {\rm{Mpc}} \ h^{-1})^{3}$, resolved with $(768)^{3}$ dark matter particles with a mass of $m_{dm} \approx 4.4 \times 10^{9} \ {\rm{M_{\odot}}} \ h^{-1}$.

As in \cite{2004A&A...416..853D}, we modified the initial conditions for the different dark energy scenarios 
adapting the initial redshift for the initial conditions in the dark energy
scenarios determined by the ratio of the linear growth factors $D_+(z)$,

\begin{equation}
  \frac{D_+(z_\mathrm{ini})}{D_+(0)}=
  \frac{D_\mathrm{+,\Lambda CDM}(z^\mathrm{ini}_\mathrm{\Lambda CDM})}
       {D_\mathrm{+,\Lambda CDM}(0)}\;.
\end{equation}

\noindent Therefore, all simulations start from the same random phases, but the amplitude of the initial fluctuations is rescaled to satisfy the constraints given by CMB.

\noindent Our reference $\Lambda$CDM model is adapted to the WMAP3 values (\cite{2007ApJS..170..377S}), with the following cosmological parameters:

\begin{itemize}
\item matter density: $\Omega_{0m}=0.268$
\item dark energy density: $\Omega_{0\Lambda}=0.732$
\item baryon density: $\Omega_{0b}=0.044$
\item Hubble parameter: $h=0.704$
\item power spectrum normalization: $\sigma_{8}=0.776$
\item spectral index: $n_{s}=0.947$
\end{itemize}

\begin{figure}
\begin{center}
\hspace*{-0.3in}
\includegraphics[width=0.5\textwidth]{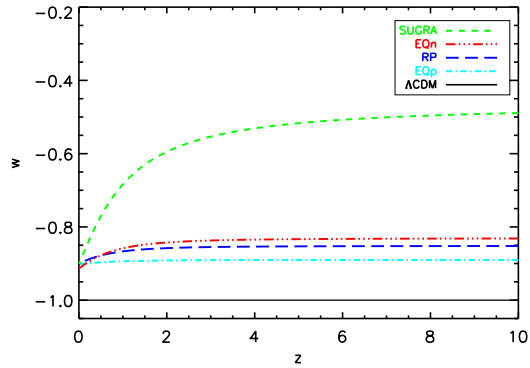}
\end{center}
\caption{Redshift evolution of the equation of state parameter $w$ for the different cosmological models considered: $\Lambda$CDM (black), RP (blue), SUGRA (green), EQp (cyan), and EQn (red).}
\label{w_z}
\end{figure}

\begin{figure}
\begin{center}
\hspace*{-0.3in}
\includegraphics[width=0.5\textwidth]{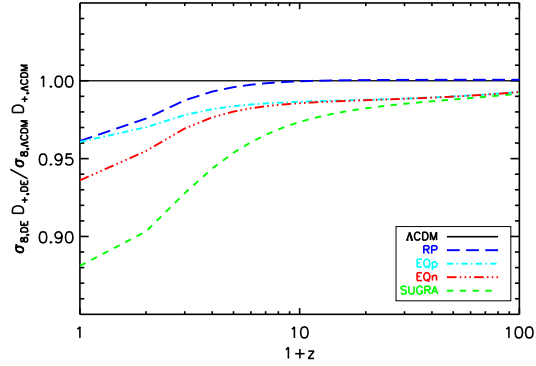}
\end{center}
\caption{Ratio between the value of $\sigma_{8} D_{+}$ for the $\Lambda$CDM (black), RP (blue), EQp (cyan), EQn (red), and SUGRA (green) cosmologies and the corresponding value for $\Lambda$CDM as a function of redshift.}
\label{sigma_8_D_ratio}
\end{figure}

We trimmed the parameters of the four dynamical dark energy models so that $w_0=w(0)\approx-0.9$ is the highest value still consistent with observational constraints in order to amplify the effects of dark energy. Fig. \ref{w_z} shows the evolution with redshift of $w$ in each cosmology. The parameters $\Omega_{0m}$, $\Omega_{0\Lambda}$, $\Omega_{0b}$, $h$, and $n_{s}$ are the same for all the models, but since we normalize all the dark energy models to CMB data from WMAP3, this leads to different values of $\sigma_{8}$ for the different cosmologies: 

\begin{equation}
\sigma_{8,\mathrm{DE}} = \sigma_{8,\mathrm{\Lambda CDM}} \frac{D_{+,\mathrm{\Lambda CDM}}(z_\mathrm{CMB})}{D_{+,\mathrm{DE}}(z_\mathrm{CMB})} \ ,
\label{sigmaDE}
\end{equation}

\noindent assuming $z_\mathrm{CMB}=1089$. This fact, along with the different evolution of the growth factor $D_{+}$ (shown in Fig. \ref{sigma_8_D_ratio}), has an impact on structure formation. Table \ref{tab} lists the parameters chosen for the different cosmological models.

\begin{table} \small
\begin{center}
\begin{tabular}{|l|c|c|c|c|}
\hline
Model & $\alpha$ & $\xi$ & $w_0$ & $\sigma_{8}$ \\ 
\hline
$\Lambda$CDM & --- & --- & $-1.0$ & $0.776$ \\
RP & $0.347$ & --- & $-0.9$ & $0.746$ \\
SUGRA & $2.259$ & --- & $-0.9$ & $0.686$ \\
EQp & $0.229$ & $+0.085$ & $-0.9$ & $0.748$ \\ 
EQn  & $0.435$ & $-0.072$ & $-0.9$ & $0.729$ \\
\hline
\end{tabular}
\end{center}
\caption{Parameters for the different cosmological models: $\alpha$ is the exponent of the inverse power-law potential; $\xi$ is the coupling in the extended quintessence models; $w_0$ is the present value of the equation of state parameter for dark energy; $\sigma_{8}$ is the normalization of the power spectrum.}
\label{tab}
\end{table}

\begin{figure}
\begin{center}
\hspace*{-0.2in}
\hbox{
\includegraphics[width=0.5\textwidth]{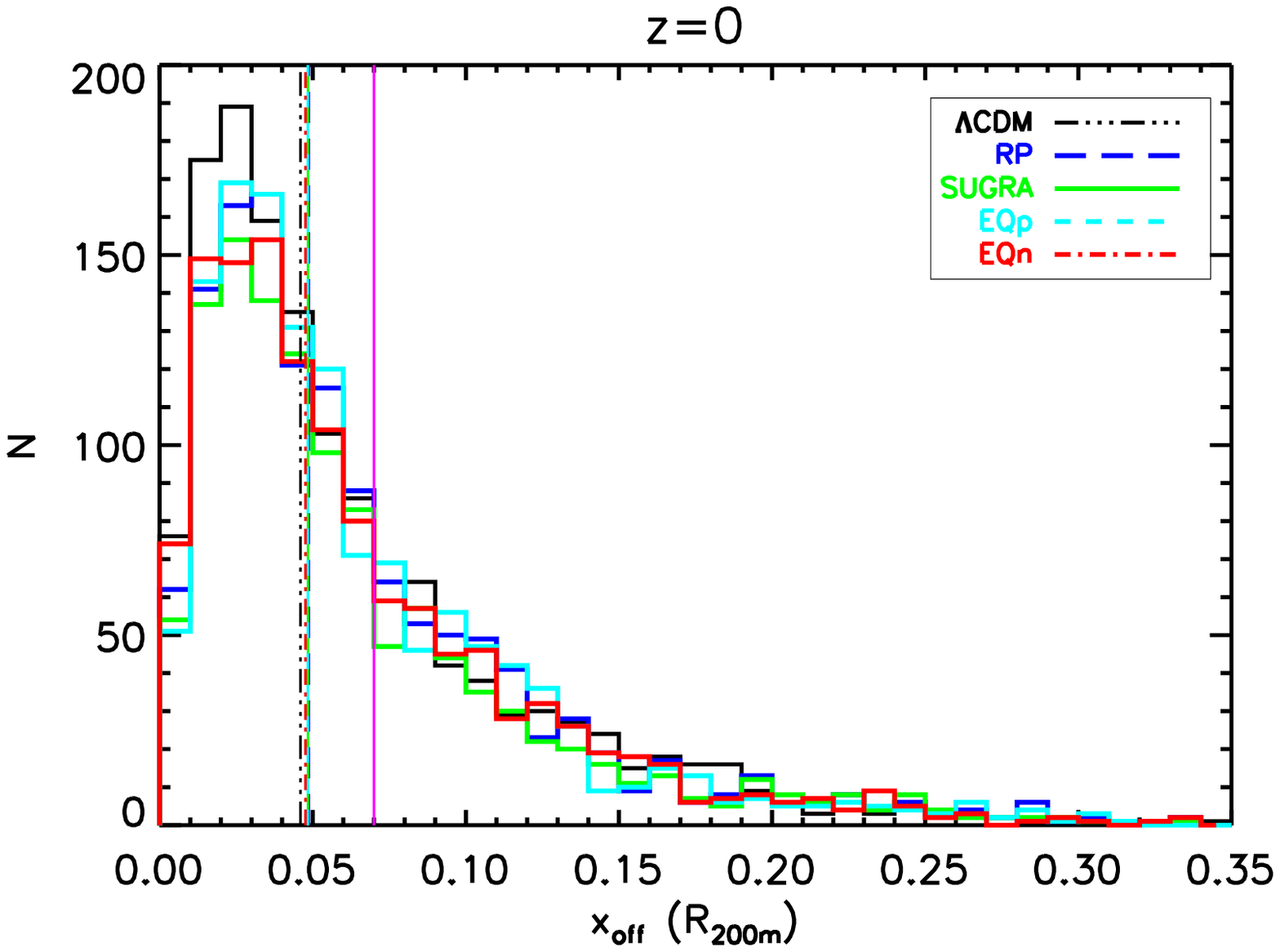}
\includegraphics[width=0.5\textwidth]{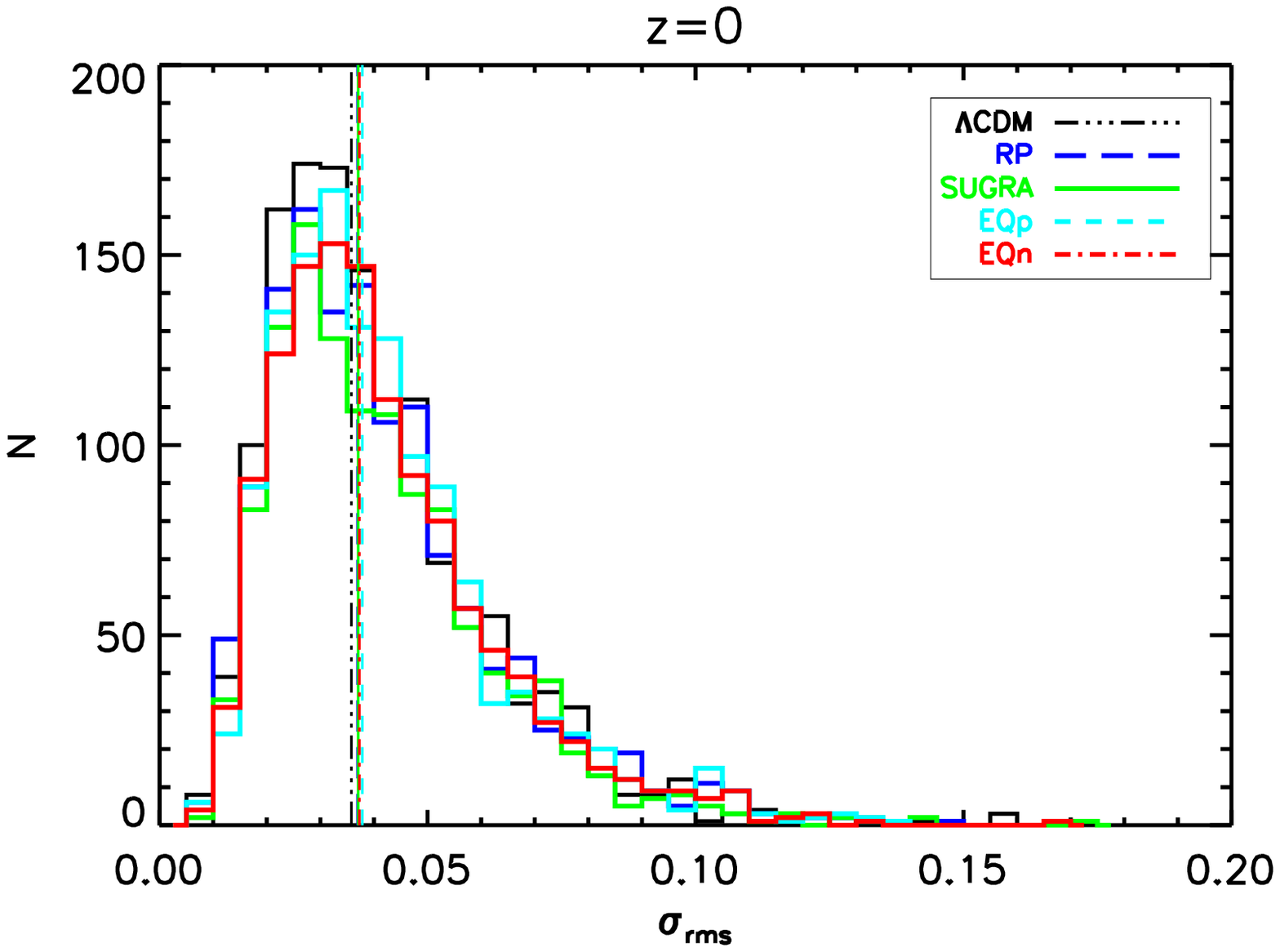}
}
\end{center}
\caption{Left-hand panel: the distribution of $x_{off}$ (in units of $R_{200m}$) for the objects in $\Lambda$CDM (black), RP (blue), SUGRA (green), EQp (cyan), and EQn (red) at $z=0$. The vertical lines of the corresponding colours mark the median value of $x_{off}$ in each cosmological model. The vertical pink line corresponds to the value defining relaxed objects, $x_{off}=0.07R_{200m}$. Right-hand panel: the same as in the left-hand panel, but for $\sigma_{rms}$.}
\label{histograms}
\end{figure}

Using the outputs of the simulations, we extract galaxy clusters from the cosmological boxes, using the spherical overdensity criterion to define the collapsed structures. We take as halo centre the position of the most bound particle. Around this particle, we construct spherical shells of matter and stop when the overdensity drops below $200$ times the {\it mean} (as opposed to {\it critical}) background density defined by $\Omega_{m}\rho_{0c}$; the radius so defined is denoted with $R_{200m}$ and the mass enclosed in it as $M_{200m}$. We consider all the halos having $M_{200m}> 10^{14} \ {\rm{M_{\odot}}} \ h^{-1}$. In addition, we selected subsamples of the $200$ objects with $M_{200m}$ closest to $7 \times 10^{13} \ {\rm{M_{\odot}}} \ h^{-1}$, $5 \times 10^{13} \ {\rm{M_{\odot}}} \ h^{-1}$, $3 \times 10^{13} \ {\rm{M_{\odot}}} \ h^{-1}$, and $10^{13} \ {\rm{M_{\odot}}} \ h^{-1}$. Starting from the centres of the halos, we construct radial profiles by binning the particles in radial bins. We concentrate on objects at $z=0$. For the following analysis, we also calculate for each cluster selected in this way the radius at which the overdensity drops below $200$ times the {\it critical} background density and denote it as $R_{200}$. The corresponding mass is indicated as $M_{200}$. It is useful to define a quantitative criterion to decide whether a cluster can be considered relaxed or not because, in general, relaxed clusters have more spherical shapes, better defined centres and thus are more representative of the self-similar behaviour of the dark matter halos. We use a simple criterion similar to the one introduced in \cite{2007MNRAS.381.1450N}: first we define $x_{off}$ as the distance between the centre of the halo (given by the most bound particle) and the barycentre of the region included in $R_{200m}$; then we define as relaxed the halos for which $x_{off}< 0.07R_{200m}$. We plot the distribution of $x_{off}$ for the objects in the five cosmological models at $z=0$ in the left-hand panel of Fig. \ref{histograms}. Note that the distribution and the median value of $x_{off}$ are similar in the different cosmological models.

\section{$c-M$ relation}

\noindent For each cluster at $z=0$ in the five cosmological models under investigation, we perform a logarithmic fit, using Poissonian errors $(\ln 10 \times \sqrt{n_{dm}})^{-1}$ (where $n_{dm}$ is the number of dark matter particles in each radial bin, of the order of $10-10^3$ depending on the mass of the object), of the three-dimensional dark matter profile $\rho_{dm}(r)$ in the region [$0.1-1$]$R_{200}$ (where the value of $R_{200}$ is taken directly from the true mass profile) with a NFW profile (\cite{1996ApJ...462..563N})

\begin{equation}
\frac{\rho_{dm}(r)}{\rho_c}=\frac{\delta}{(r/r_{s})(1+r/r_{s})^{2}} \ ,
\label{NFW_c-M}
\end{equation}

\noindent where $\rho_{c}$ is the critical density, $r_{s}$ is the scale radius and $\delta$ is a characteristic density contrast. Then, instead of defining $c_{200} \equiv R_{200}/r_{s}$, we directly find the concentration parameter $c_{200}$ from the normalization of the NFW profile

\begin{equation}
\delta=\frac{200}{3}\frac{c_{200}^3}{\left[\ln (1+ c_{200}) - \frac{c_{200}}{1+c_{200}}\right]} \ .
\label{c_fit}
\end{equation}

\noindent We require the central density parameter $\delta$ to be greater than $100$ and the scale radius $r_{s}$ to be within [$0.1-1$]$R_{200}$. We exclude the inner regions from the fit because we are limited in resolution inside a given radius. We indicate the dark matter concentration found in this way as $c_{200dm}$. We define the rms deviation $\sigma_{rms}$ as

\begin{equation}
\sigma^2_{rms} = \frac{1}{N_{bins}} \sum_{i=1}^{N_{bins}} [{\rm{log_{10}}} \rho_{dm_{i}} - {\rm{log_{10}}} \rho_{NFW_{i}}]^2 \ ,
\label{sigma}
\end{equation}

\noindent where $N_{bins}$ is the number of radial bins over which the fit is performed and $\rho_{NFW}$ is the best-fitting NFW profile. We plot the distribution of $\sigma_{rms}$ for the objects in the five cosmological models at $z=0$ in the right-ended panel of Fig. \ref{histograms}. Note that the distribution and the median value of $\sigma_{rms}$ are similar in the different cosmological models, meaning that the NFW profile is as good as in $\Lambda$CDM in describing the dark matter profile of galaxy clusters in dark energy cosmologies.

We bin the objects in the complete sample in groups of 200, so that we have bins around $10^{13} \ {\rm{M_{\odot}}} \ h^{-1}$, $3 \times 10^{13} \ {\rm{M_{\odot}}} \ h^{-1}$, $5 \times 10^{13} \ {\rm{M_{\odot}}} \ h^{-1}$, and $7 \times 10^{13} \ {\rm{M_{\odot}}} \ h^{-1}$. For halos more massive than $10^{14} \ {\rm{M_{\odot}}} \ h^{-1}$, we bin the objects starting from the low-mass ones, so that the most massive bin can contain less than 200 objects. The analysis for the relaxed sample is done by selecting the relaxed objects inside each bin. Once we have $c_{200dm}$ for each object in each mass bin, since the distribution of $c_{200dm}$ is log-normal inside each bin, we evaluate the mean $M_{200}$ and the mean and rms deviation of ${\rm{log_{10}}} c_{200dm}$ in each bin, for the two samples. In the following of the proceeding, when we indicate the value of $c_{200dm}$ in a mass bin, we refer to $10^{\langle{{\rm{log_{10}}} c_{200dm}}\rangle}$.

With the mean and rms deviation of ${\rm{log_{10}}} c_{200dm}$ in each bin at hand, we fit, for the complete and relaxed samples, the binned $c-M$ relation using

\begin{equation}
{\rm{log_{10}}} c_{200}={\rm{log_{10}}} A + B \ {\rm{log_{10}}} \left( \frac{M_{200}}{10^{14} \ {\rm{M_{\odot}}}} \right) \ ,
\label{c-M}
\end{equation}

\noindent where ${\rm{log_{10}}} c_{200}$ and $M_{200}$ are the mean values in each bin. For the error on the mean of ${\rm{log_{10}}} c_{200dm}$ in each bin, $\sigma_{\bar{c}}$, we use the rms deviation of ${\rm{log_{10}}} c_{200dm}$ divided by the square root of the number of objects in the bin. For each fit we also define

\begin{equation}
\chi^2 = \sum_{j=1}^{N_{mass}} \left( \frac{{\rm{log_{10}}}  c_{200dm_{j}} - {\rm{log_{10}}}  c_{200fit_{j}}}{\sigma_{\bar{c}_{j}}} \right)^2 \ ,
\end{equation}

\noindent where $N_{mass}$ is the number of mass bins over which the fit is performed and $c_{200fit}$ is obtained from the best fit of equation (\ref{c-M}), and evaluate the reduced chi-squared $\tilde{\chi}^2$, {\it{i.e.}} ${\chi}^2$ divided by the number of degrees of freedom.

In the reference $\Lambda$CDM model, relaxed objects have a higher normalization and a shallower slope with respect to the complete sample (see Table \ref{tab_parameters_core_cut_dm} for details). By comparing our results for $\Lambda$CDM with previous works in literature (see \cite{2013MNRAS.428.2921D} for the comparison), we find a good agreement, in particular when the values of the cosmological parameters are similar, as in \cite{2008MNRAS.391.1940M}. Thus, when comparing the impact of different dark energy models on the $c-M$ relation, we can rely on our $\Lambda$CDM model as a reference. The $c-M$ relation for galaxy clusters extracted from dark matter only simulations of different dark energy models, including RP and SUGRA, has been studied in \cite{2004A&A...416..853D}. They fit a formula similar to equation (\ref{c-M}) and find that, when the same $\sigma_{8}$ is used for all the models, the normalization of the $c-M$ relation for dark energy cosmologies is higher compared to $\Lambda$CDM, depending on the ratio between the growth factors through

\begin{equation}
A_\mathrm{DE} \rightarrow A_\mathrm{\Lambda CDM} \frac{D_\mathrm{+,DE}(z_{coll})}{D_\mathrm{+,\Lambda CDM}(z_{coll})} \ ,
\label{c_dark_energy}
\end{equation}

\noindent where the collapse redshifts $z_{coll}$ are evaluated following the prescriptions of \cite{2001ApJ...554..114E}. When $\sigma_{8}$ values are normalized to CMB data, as we do in this work, the normalization of the $c-M$ relation for dark energy cosmologies is lower compared to $\Lambda$CDM. We find that, in order to recover the values of the normalization they quote in this case, equation (\ref{c_dark_energy}) should be multiplied by the ratio between the values of $\sigma_{8}$, {\it{i.e.}} $\sigma_{8,\mathrm{DE}} / \sigma_\mathrm{8,\mathrm{\Lambda CDM}}$. This fact goes in the same direction as what found in \cite{2008MNRAS.391.1940M}, where models with higher $\sigma_{8}$ also have a higher normalization of the $c-M$ relation.

\section{Results}

\noindent We compare the $c-M$ relation for the dark energy models under investigation with the one derived for the $\Lambda$CDM cosmology. In Table \ref{tab_parameters_core_cut_dm} we summarize the best-fitting parameters, the standard errors and the reduced chi-squared of the $c-M$ relation equation (\ref{c-M}) for the five cosmological models here considered, both for the complete and relaxed samples. For the complete sample, the differences in the normalization $A$ between $\Lambda$CDM and the other cosmological models are less than $10\%$, with EQn being the only model having a higher normalization. The slope $B$ is within $5\%$ of the $\Lambda$CDM value for all the models with the exception of EQn, which shows a $30\%$ flatter slope. For the slope the differences among the models, excluding EQn, are smaller than the standard errors, while for the normalization these differences are significant. If we limit ourselves to the best-fitting values, given that the slope is almost identical and that all the cosmological parameters except $\sigma_{8}$ are fixed, we expect that the normalization should follow the values of $\sigma_{8}$, {\it{i.e.}} the higher $\sigma_{8}$ the higher the normalization (see \cite{2008MNRAS.391.1940M}), and $D_{+}$, {\it{i.e.}} the higher $D_{+}$ at $z_{coll}$ the higher the normalization (see \cite{2004A&A...416..853D}). The quantity controlling the normalization is thus expected to be $\sigma_{8} D_{+}(z_{coll})$, which is plotted as a function of redshift in Fig. \ref{sigma_8_D_ratio} for the five cosmological models. Independently of the precise definition of $z_{coll}$, the cosmological model with the highest value of this quantity is $\Lambda$CDM, followed by RP, EQp, EQn, and SUGRA. We do expect the normalization of the $c-M$ relation to follow the same order, with $\Lambda$CDM having the highest and SUGRA the lowest. Instead we see that, on the one hand, EQp which has the third highest $\sigma_{8} D_{+}$ has the lowest normalization while, on the other hand, EQn which has the second lowest $\sigma_{8} D_{+}$ has the highest normalization. The relative order of $\sigma_{8} D_{+}$ and $A$ is preserved for $\Lambda$CDM, RP and SUGRA, as in \cite{2004A&A...416..853D}. 

\begin{table} \small
\begin{center}
\begin{tabular}{|cc|cc|cc|c|cc|cc|c|}
\hline Model & $\sigma_8$ & A & $\sigma_{A}$ & $B$ & $\sigma_{B}$ & $\tilde{\chi}^2$ & A & $\sigma_{A}$ & $B$ & $\sigma_{B}$ & $\tilde{\chi}^2$ \\
\hline & & \multicolumn{5}{|c|}{all} & \multicolumn{5}{|c|}{relaxed} \\
\hline
$\Lambda$CDM & $0.776$ & $3.59$ & $0.05$ & $-0.099$ & $0.011$ & $0.48$ & $4.09$ & $0.05$ & $-0.092$ & $0.011$ & $0.66$ \\
RP & $0.746$ & $3.54$ & $0.05$ & $-0.103$ & $0.011$ & $1.14$ & $4.08$ & $0.05$ & $-0.081$ & $0.011$ & $0.92$ \\
SUGRA & $0.686$ & $3.41$ & $0.05$ & $-0.098$ & $0.013$ & $1.50$ & $3.94$ & $0.06$ & $-0.081$ & $0.012$ & $1.55$ \\
EQp & $0.748$ & $3.36$ & $0.05$ & $-0.097$ & $0.012$ & $0.35$ & $3.84$ & $0.05$ & $-0.097$ & $0.011$ & $1.32$ \\
EQn & $0.726$ & $3.70$ & $0.05$ & $-0.069$ & $0.013$ & $0.78$ & $4.25$ & $0.06$ & $-0.081$ & $0.013$ & $0.51$ \\
\hline
\end{tabular}
\end{center}
\caption{Best-fitting parameters, standard errors and reduced chi-squared $\tilde{\chi}^2$ of the $c-M$ relation equation (4.4) for dark matter density profile fit in the region [$0.1-1$]$R_{200}$ for the complete and relaxed samples of the five different cosmological models at $z=0$.}
\label{tab_parameters_core_cut_dm}
\end{table}

\noindent For the relaxed sample, compared to $\Lambda$CDM, the differences in the normalization are less than $10\%$, while the differences in the slope can almost reach $15\%$, but they are compatible with the standard errors. Also in this case, the most extreme cosmologies are EQp and EQn, whose normalization goes in the opposite direction with respect to their $\sigma_{8} D_{+}$. This fact confirms the conclusions we have drawn from the complete sample. The values of the reduced chi-squared indicate that equation (\ref{c-M}) is a good parametrization of the $c-M$ relation for almost all cosmological models. Only SUGRA has high values both for the complete and relaxed samples.

\noindent Our results are in good qualitative agreement with the findings of \cite{2011ApJ...728..109L}, where halos in extended quintessence models have lower (higher) concentrations with respect to the $\Lambda$CDM case for positive (negative) values of the scalar field coupling.

\begin{figure*}
\begin{center}
\hspace*{-0.2in}
\hbox{
\includegraphics[width=0.5\textwidth]{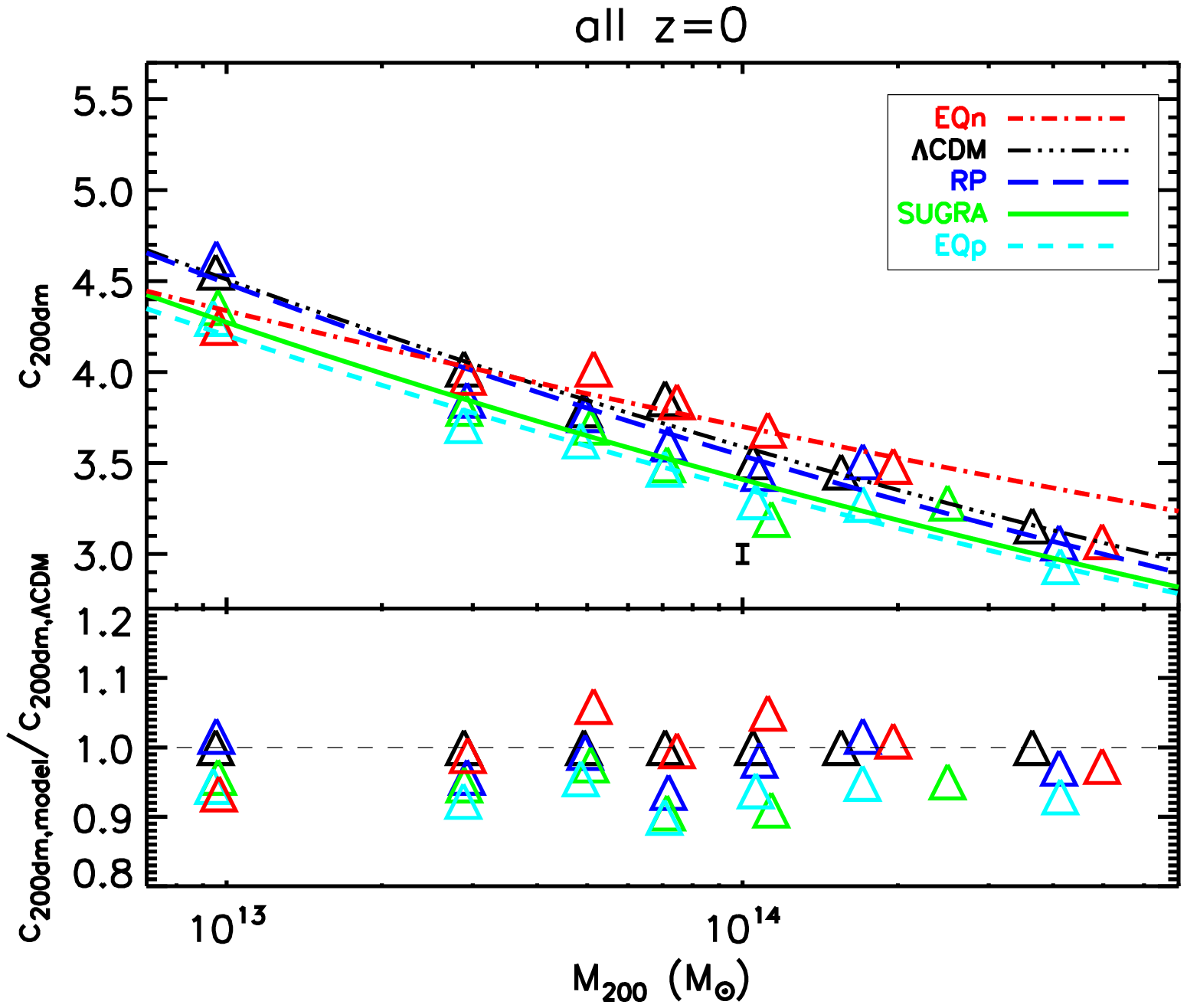}
\includegraphics[width=0.5\textwidth]{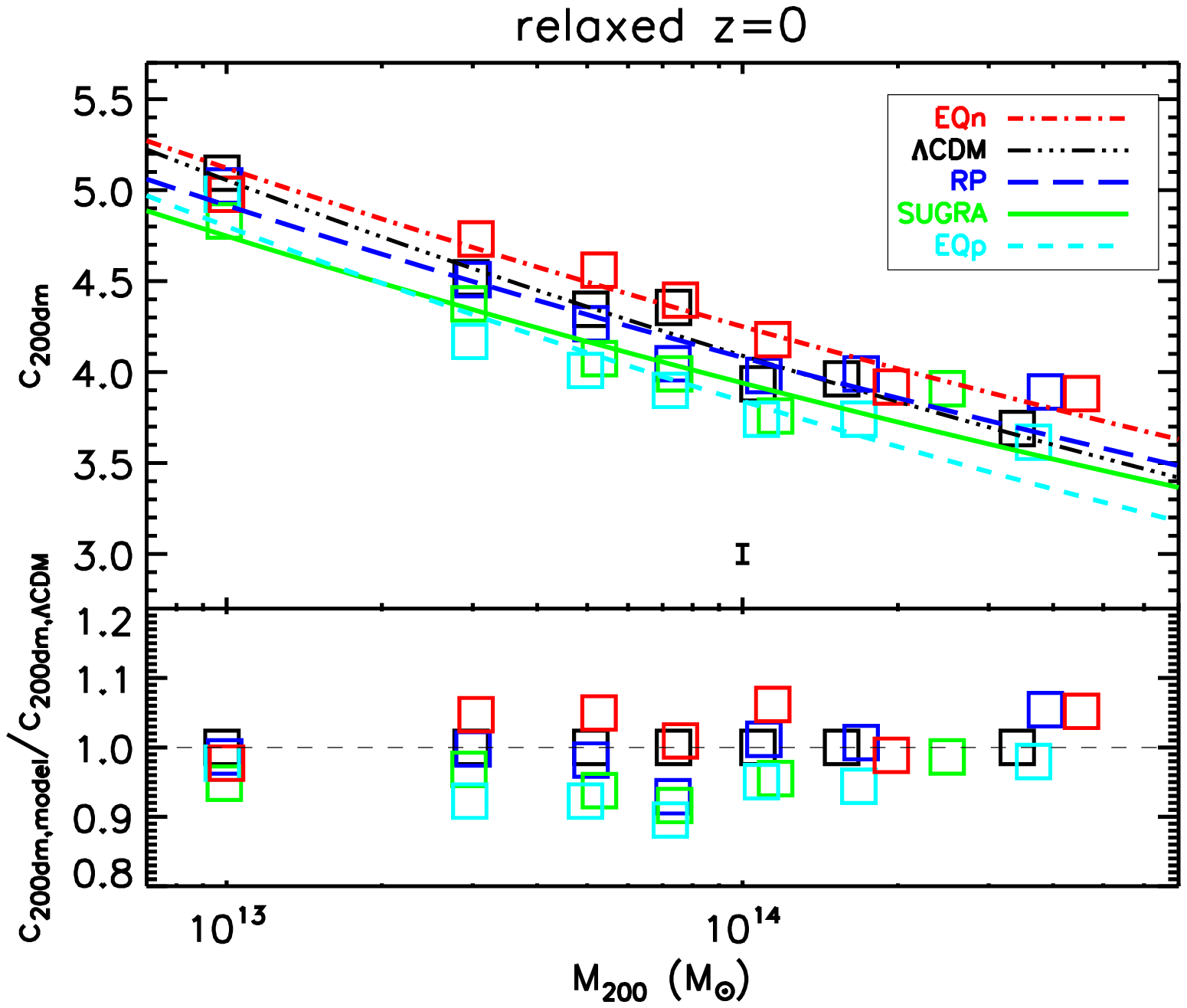}
}
\end{center}
\caption{Left-hand panel: the values of $c_{200dm}$ for the complete sample of the $\Lambda$CDM (black), RP (blue), SUGRA (green), EQp (cyan), and EQn (red) cosmologies at $z=0$. The lines of the corresponding colours are our best fit of $c-M$ relation equation (4.4) and the vertical black bar is the error on the normalization of $\Lambda$CDM as listed in Table 2. The symbols in the low part of the panel are the ratios between $c_{200dm}$ for the model and $c_{200dm}$ for $\Lambda$CDM. Right-hand panel: the same as in the left-hand panel, but for the relaxed sample.}
\label{cdm-M_fit}
\end{figure*}

We plot the best-fitting $c-M$ relation for all the cosmological models, along with the binned data, in Fig. \ref{cdm-M_fit}. We clearly see that the results on the normalization are due to differences in the concentrations over a wide mass range. If we look, for example, at the complete sample (left-hand panel of Fig. \ref{cdm-M_fit}), we see that the different slope of EQn is mainly originated by the less massive bin. But with the exception of this bin, EQn shows the highest concentration in almost all the mass bins, while in general EQp has the lowest concentration. For the relaxed sample (right-hand panel of Fig. \ref{cdm-M_fit}), the relative behaviour of the different cosmological models is even clearer, and indeed the differences in the slope are less pronounced. 

\begin{figure}
\begin{center}
\hspace*{-0.3in}
\hbox{
\includegraphics[width=0.5\textwidth]{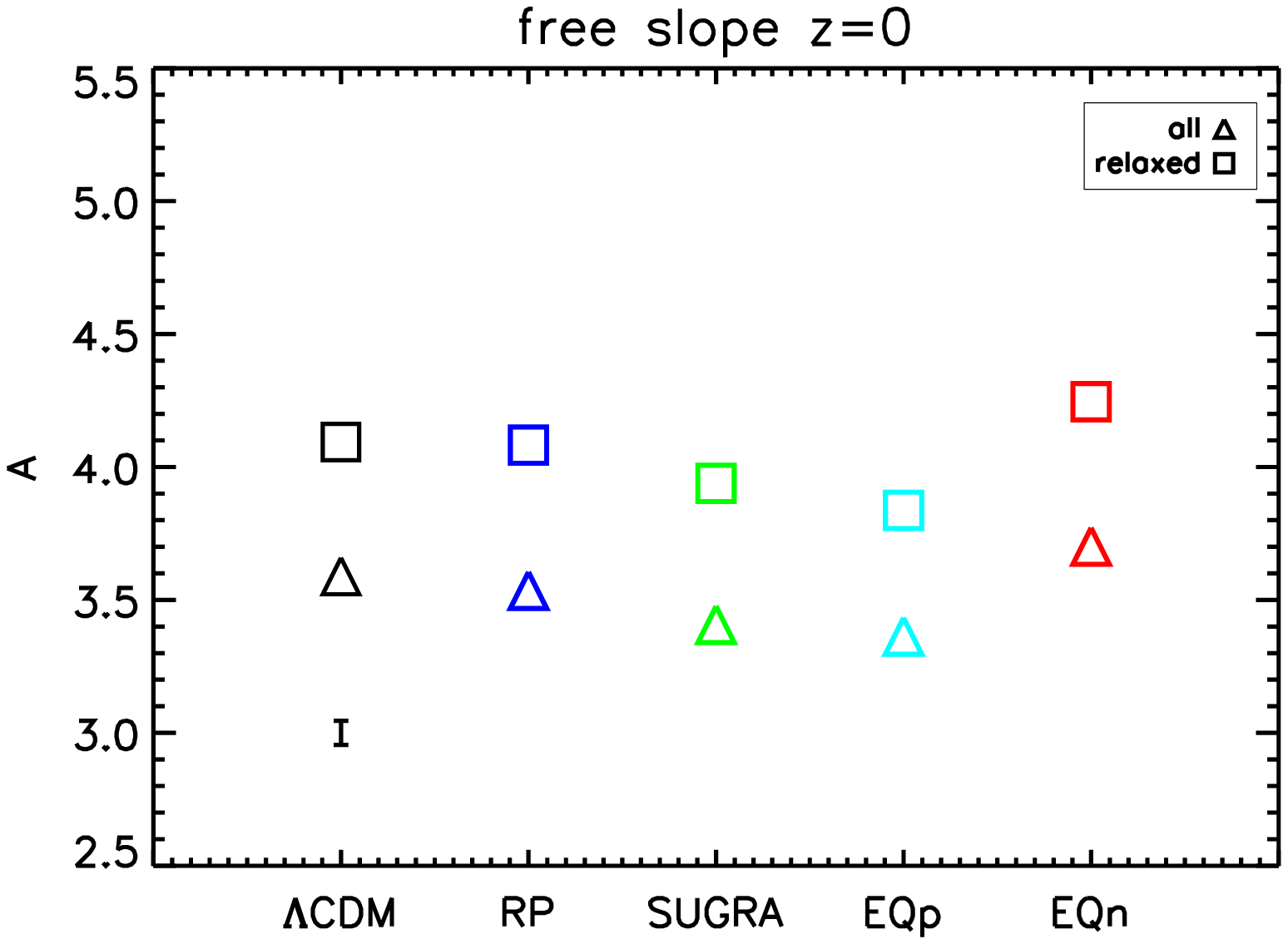}
\includegraphics[width=0.5\textwidth]{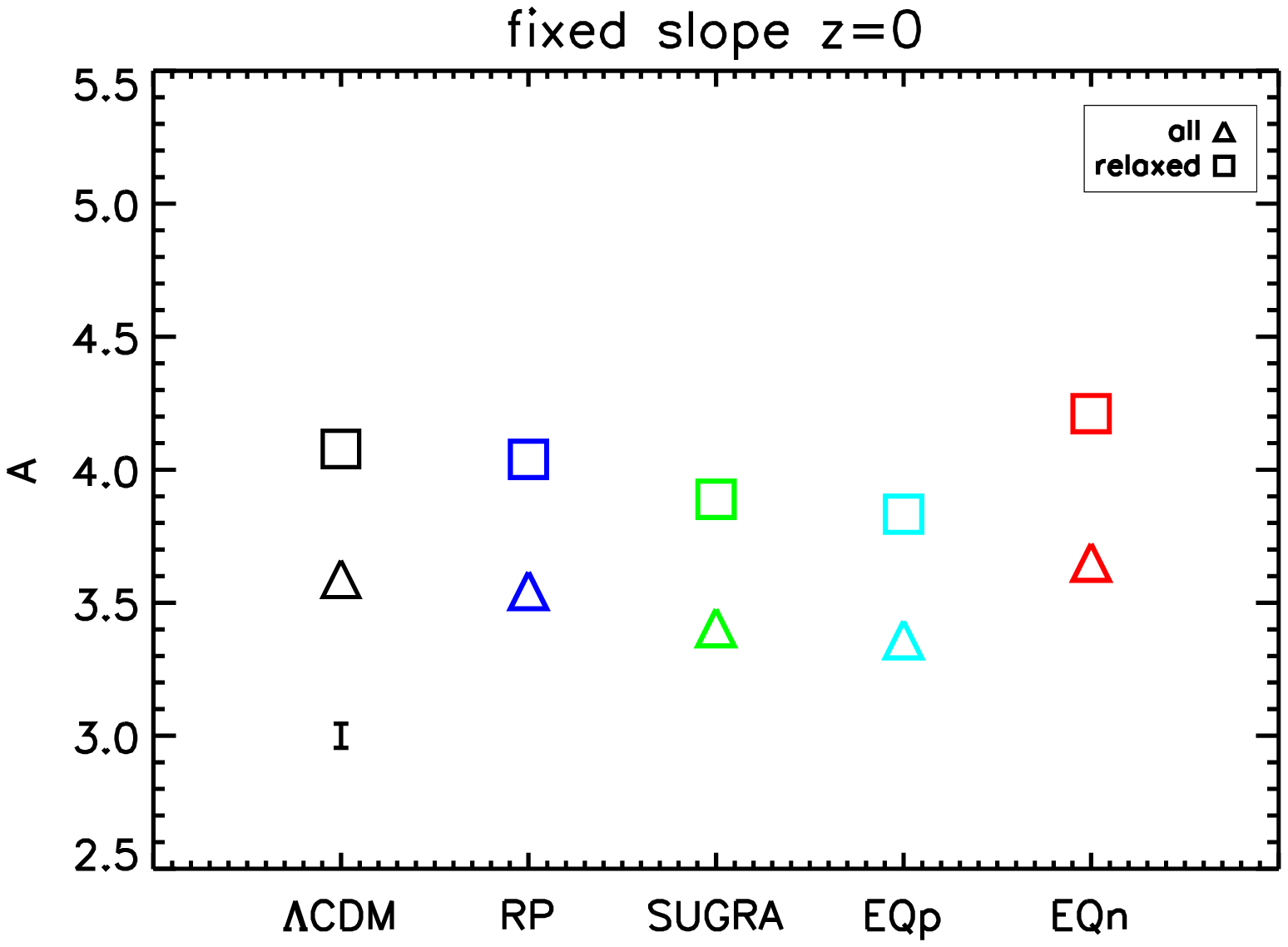}
}
\end{center}
\caption{Left-hand panel: best-fitting normalization comparison for equation (4.4) for the $\Lambda$CDM (black), RP (blue), SUGRA (green), EQp (cyan), and EQn (red) cosmologies. Triangles: dark matter profile fit, complete sample. Squares: dark matter profile fit, relaxed sample. The vertical black bar is the error on the normalization of the complete sample of $\Lambda$CDM. Right-hand panel: the same as left-hand panel but with $B$ fixed at the best-fitting value for the complete sample of $\Lambda$CDM at $z=0$.}
\label{normalization}
\end{figure}

Before drawing our conclusions about the EQ models, we want to take into account the dependence of the normalization on the slope that characterizes the $c-M$ relation in the different cosmological models. To do this, we fix the slope at the best-fitting value for the complete sample of $\Lambda$CDM at $z=0$ ({\it{i.e.}} $B=-0.099$, see Table \ref{tab_parameters_core_cut_dm}) and we fit equation (\ref{c-M}) with only $A$ as a free parameter. We report the results in Table \ref{tab_normalization} and plot them in Fig. \ref{normalization}, which summarizes almost all the information on the $c-M$ relation at $z=0$ for the cosmological models under investigation. We show the values of the reduced chi-squared of the fit as a reference, but we do not discuss them because we are imposing the slope for $\Lambda$CDM also to other models. Also in this case, relaxed objects have a higher normalization compared to the complete sample. Then, as a general trend, both fixing or keeping the slope free, the normalization is decreasing going from $\Lambda$CDM to RP to SUGRA, independently of the dynamical state. Finally EQn always has the highest normalization while EQp alway has the lowest. The behaviour of $\Lambda$CDM, RP and SUGRA is in agreement with the simple idea that the normalization of the $c-M$ relation is driven by the value of $\sigma_{8} D_{+}$, but the one of EQp and EQn is not. 

\noindent We hint that the behaviour of EQp and EQn is linked to the redshift evolution of the effective gravitational interaction $\tilde{G}$, as pointed out in Section \ref{models}. In fact, in contrast with $\Lambda$CDM, RP and SUGRA, in EQ models the gravitational constant $G$ is substituted by $\tilde{G}$, which is higher (lower) than $G$ at high redshift for positive (negative) values of the coupling constant $\xi$, while it is equal to $G$ at $z=0$ in order to recover General Relativity (see Fig. \ref{dG_z}). This means that in EQp gravity becomes weaker at low redshift compared to high redshift, while in EQn it becomes stronger. Thus one can expect that in EQp (EQn) the halos expand (contract) due to the change in the gravitational interaction, resulting in lower (higher) concentrations with respect to the case in which gravity is constant.

\begin{table*} \small
\begin{center}
\begin{tabular}{|cc|cc|cc|c|cc|cc|c|}
\hline Model & $\sigma_8$ & A & $\sigma_{A}$ & $B$ & $\sigma_{B}$ & $\tilde{\chi}^2$ & A & $\sigma_{A}$ & $B$ & $\sigma_{B}$ & $\tilde{\chi}^2$ \\
\hline & & \multicolumn{5}{|c|}{all} & \multicolumn{5}{|c|}{relaxed} \\
\hline
$\Lambda$CDM & $0.776$ & $3.59$ & $0.05$ & $-0.99$ & $0.11$ & $0.48$ & $4.08$ & $0.04$ & $-0.99$ & --- & $0.62$ \\
RP & $0.746$ & $3.55$ & $0.04$ & $-0.99$ & --- & $0.97$ & $4.04$ & $0.05$ & $-0.99$ & --- & $1.19$ \\
SUGRA & $0.686$ & $3.41$ & $0.04$ & $-0.99$ & --- & $1.20$ & $3.89$ & $0.04$ & $-0.99$ & --- & $1.69$ \\
EQp & $0.748$ & $3.36$ & $0.04$ & $-0.99$ & --- & $0.30$ & $3.83$ & $0.04$ & $-0.99$ & --- & $1.11$ \\
EQn & $0.726$ & $3.65$ & $0.04$ & $-0.99$ & --- & $1.56$ & $4.21$ & $0.05$ & $-0.99$ & --- & $0.74$ \\
\hline
\end{tabular}
\end{center}
\caption{Best-fitting parameters, standard errors and reduced chi-squared $\tilde{\chi}^2$ of the $c-M$ relation equation (4.4), with $B$ fixed at the best-fitting value for the complete sample of $\Lambda$CDM at $z=0$, for dark matter density profile fit in the region [$0.1-1$]$R_{200}$ for the complete and relaxed samples of the five different cosmological models at $z=0$.}
\label{tab_normalization}
\end{table*}

\section{Conclusions}

\noindent In this proceeding, we reviewed the $c-M$ relation for the halos extracted from the simulation set introduced in \cite{2011MNRAS.415.2758D} and \cite{2013MNRAS.428.2921D}. We find that the normalization of the $c-M$ relation in dynamical dark energy cosmologies is different with respect to the $\Lambda$CDM one, while the slope is more compatible. In particular, at $z=0$, the differences in the normalization for RP and SUGRA when compared to $\Lambda$CDM reflect the differences in $\sigma_{8}D_{+}$, with models having a higher $\sigma_{8}D_{+}$ also having a higher normalization. This simple scheme is not valid for the EQp and EQn scenarios. In the former case, the normalization is lower than expected considering $\sigma_{8} D_{+}$, while in the latter it is higher, and indeed EQn is always the model with the highest normalization, regardless of the dynamical state of the objects. This behaviour is due to the different redshift evolution of the effective gravitational interaction $\tilde{G}$ that characterizes these models. Indeed, going from high to low redshift, $\tilde{G}$ decreases (increases) for EQp (EQn), making the halos expanding (contracting) and thus decreasing (increasing) the concentration. This is a very important result because it shows a direct manifestation of the coupling between gravity and the quintessence scalar field that cannot be seen at the background level but can be investigated in the non-linear regime.

\section*{Acknowledgments}

\noindent Computations have been performed at the ``Leibniz-Rechenzentrum'' with CPU time assigned to the Project ``h0073''.
We acknowledge financial contributions from contracts ASI I/016/07/0 COFIS, ASI-INAF I/023/05/0, ASI-INAF I/088/06/0, ASI
`EUCLID-DUNE' I/064/08/0, PRIN MIUR 2008 ``Dark energy and cosmology with large galaxy survey'', and PRIN INAF 2009 ``Towards an Italian network of computational cosmology''. We thank Matthias Bartelmann, Andrea Macci\`o and Marco Baldi for useful discussions.

\end{document}